\documentclass[aps,prl,twocolumn,groupedaddress, floatfix,showpacs]{revtex4}

\usepackage{amsmath} 
\usepackage{amssymb} 

\usepackage{graphicx}

\newcommand{\vek}    [1] {\textrm{\textbf{{#1}}}} 
\newcommand{\hatb}   [1] {\hat{\textrm{\textbf{#1}}}} 
\newcommand{\bk}     [2] {\langle #1 | #2 \rangle}
\newcommand{\bsymbol}[1] {\mbox{\boldmath$\displaystyle#1$\unboldmath}}

\begin{document}


\title{Alignment-Dependent Ionization of
       Molecular Hydrogen in Intense Laser Fields}

\author{Yulian V.~Vanne}

\author{Alejandro Saenz}

\affiliation{
     AG Moderne Optik,
     Institut f\"ur Physik,
     Humboldt-Universit\"at zu Berlin,
     Hausvogteiplatz 5-7, D\,-\,10\,117 Berlin, Germany
}

\date{\today}

\begin{abstract}
     The alignment dependence of the ionization behavior of H$_2$ 
     exposed to intense ultrashort laser pulses is investigated 
     on the basis of solutions of the full time-dependent Schr\"odinger 
     equation within the fixed-nuclei and dipole approximation. 
     The total ionization yields as well as the energy-resolved 
     electron spectra have been calculated for a parallel and a 
     perpendicular orientation of the molecular axis with respect 
     to the polarization axis of linear polarized laser pulses. 
     For most, but not all considered laser peak intensities 
     the parallel aligned molecules are easier to ionize. 
     Furthermore, it is shown that the velocity formulation of the 
     strong-field approximation predicts a simple interference 
     pattern for the ratio of the energy-resolved electron spectra 
     obtained for the two orientations, but this is not confirmed 
     by the full ab initio results. 
\end{abstract}

\pacs{32.80.Rm, 33.80.Rv}

\maketitle

%
Time-resolved imaging of the dynamics of nuclei and electrons on a 
femtosecond or even sub-femtosecond time scale is a prerequisite for 
the real-time investigation of the formation and breaking of 
chemical bonds. Ultrashort laser pulses have recently been demonstrated 
to pave a possible path to the experimental realization of this 
long-standing dream. For example, ways have been proposed and 
experimentally verified that allow monitoring nuclear motion with 
sub-femtosecond and sub-{\AA}ngstrom resolution in real time 
\cite{sfm:bake06,sfm:goll06,sfm:ergl06}. It was 
also experimentally demonstrated that the high-harmonic radiation or 
the electrons emitted in an intense laser pulse may in principle reveal 
information on the electronic structure \cite{sfm:itat04,sfm:meck08} 
and thus have the potential for time-resolved imaging of changes of the 
electronic structure in, e.\,g., a chemical reaction. To reach this goal 
it is, however, important to understand the relation between electronic 
structure and the strong-field response of molecules. This includes the 
fundamental question whether the rather clear correspondence between 
the symmetry of the highest-occupied molecular orbital (HOMO) and the 
strong-field signal as indicated for N$_2$ and O$_2$  in 
\cite{sfm:itat04,sfm:meck08} is really a universal phenomenon. 

Already some time ago it was found that within the molecular strong-field 
approximation (MO-SFA) --- formulated in velocity gauge (VG) and within the 
framework of a linear combination of atomic orbitals (LCAO) --- the 
molecular response to intense laser fields should reflect the symmetry 
of the highest occupied molecular orbital (HOMO) 
\cite{sfm:tale98b,sfm:muth00}. 
In the case of diatomic molecules, a simple interference picture 
arises in the MO-SFA-VG that seems to plausibly explain the occurrence 
or absence of {\it suppressed ionization} \cite{sfm:muth00}. The term  
suppressed ionization describes the effect that a molecule with the same 
ionization potential as the one of some so-called companion atom is 
harder to ionize in an intense laser pulse. For example, 
molecular nitrogen shows a similar ionization behavior as atomic 
Ar, while the ionization yield of oxygen is much smaller than the one 
of Xe atoms; despite the almost identical ionization potentials of 
either N$_2$ and Ar or O$_2$ and Xe. Although suppressed ionization 
is also predicted by tunneling models like molecular 
Ammosov-Delone-Krainov (MO-ADK) \cite{sfm:tong02} or the length-gauge 
formulation of the MO-SFA \cite{sfm:kjel05a}, these theories do not 
provide a simple interference picture for the effect. On the other 
hand, the energy-resolved electron spectra measured in \cite{sfm:gras01} 
seemed to further support the concept of of symmetry-induced 
quantum-interference effects as predicted by MO-SFA-VG.    

One key step towards time-resolved imaging is the measurement of 
the molecular strong-field response within a molecule-fixed 
coordinate system. This is also the basic difference between the 
electron spectra measured in \cite{sfm:meck08} in comparison to 
the ones in \cite{sfm:gras01}. While the former are obtained as 
a function of the alignment between the molecular axis and the 
laser-field axis, the latter are averaged over all orientations. 
This explain the increased recent interest even in alignment-dependent 
total ionization yields of molecules in intense laser pulses 
\cite{sfm:pavi07,sfm:stau09,sfm:magr09}. In fact, the results in 
\cite{sfm:pavi07} seem to indicate that, at least for N$_2$ and 
O$_2$, structural information (on the HOMO) may be obtained 
already from such integral, but angular-resolved ion yields. 
Besides the problem that the results obtained in~\cite{sfm:pavi07} 
for CO$_2$ seem to be difficult to interpret as a simple mapping 
of its HOMO, already relatively simple diatomic molecules like 
O$_2$ and N$_2$ possess a HOMO that is not necessarily well described 
as a linear combination of two atomic orbitals, as is easily seen 
from the recent debate about the prediction of MO-SFA-VG for 
the parallel to perpendicular strong-field ionization yield of 
N$_2$ (see~\cite{sfm:usac06} and references therein). Therefore, 
molecular hydrogen with a comparatively simple orbital structure and 
the lack of core orbitals that may disturb the strong-field response   
appears to be a perfect candidate to investigate whether the 
interference effects predicted by MO-SFA-VG occur. Furthermore, 
H$_2$ has at least for fixed nuclei now become accessible to 
in principle exact theoretical calculations, even for a 
non-parallel orientation of the molecular axis with respect 
to the field that requires a full 6-dimensional 
treatment~\cite{sfm:vann08}. Its extension to laser fields with 
a wavelength of 800\,nm and the extraction of energy-resolved 
electron spectra is reported in this work.  

In fact, the purpose of this Letter is threefold. First, the alignment 
dependence of the total ionization yield of H$_2$ in intense laser pulses 
with a wavelength of about 800\,nm (Ti:sapphire) is investigated 
by means of a full solution of the time-dependent Schr\"odinger equation 
(TDSE) describing both correlated electrons in full dimensionality within 
the fixed-nuclei and the non-relativistic dipole approximations. It is 
shown that the ratio of ionization yields for parallel and perpendicular 
alignment is a non-monotonic function as a function of laser peak intensity, 
even if focal-volume averaging is performed, and is in reasonable 
agreement to a recent experiment~\cite{sfm:stau09}. 
Second, it is demonstrated 
theoretically that MO-SFA-VG predicts a clear interference pattern 
in the ratio of energy-resolved electron spectra for a parallel 
and a perpendicular alignment of a molecule like H$_2$. 
Finally, the corresponding energy-resolved electron spectra are 
extracted from the TDSE calculation and it is demonstrated that the 
simple interference pattern predicted by MO-SFA-VG is not confirmed.

\begin{figure}
\begin{center}
     \includegraphics[width=0.44\textwidth]{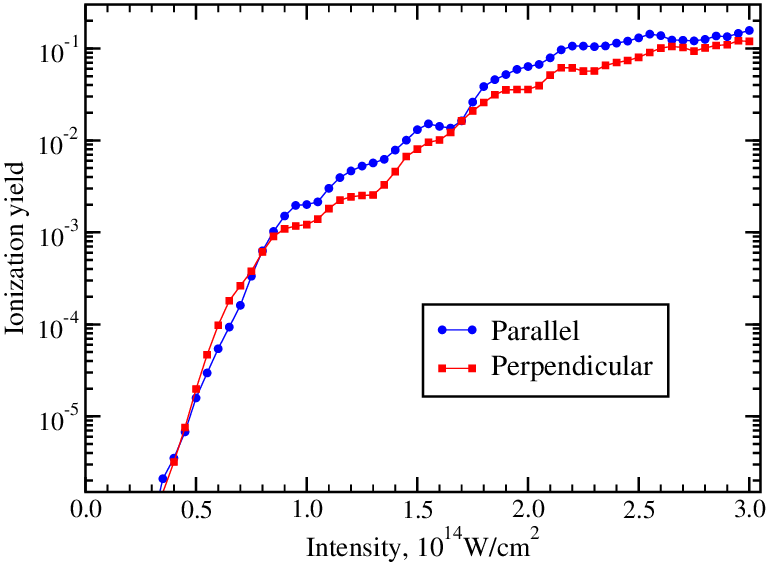}
     \caption{\label{fig:Comp_IonY}
     (Color online) Ionization yields of
     H$_2$ (internuclear separation $R=1.4\,a_0$) 
     for a 10-cycle cos$^2$-shaped 800\,nm laser pulse and a 
     parallel (blue circles) or perpendicular (red squares) 
     alignment with respect to the laser polarization vector.}
\end{center}
\end{figure}

\begin{figure} 
     \includegraphics[width=0.44\textwidth]{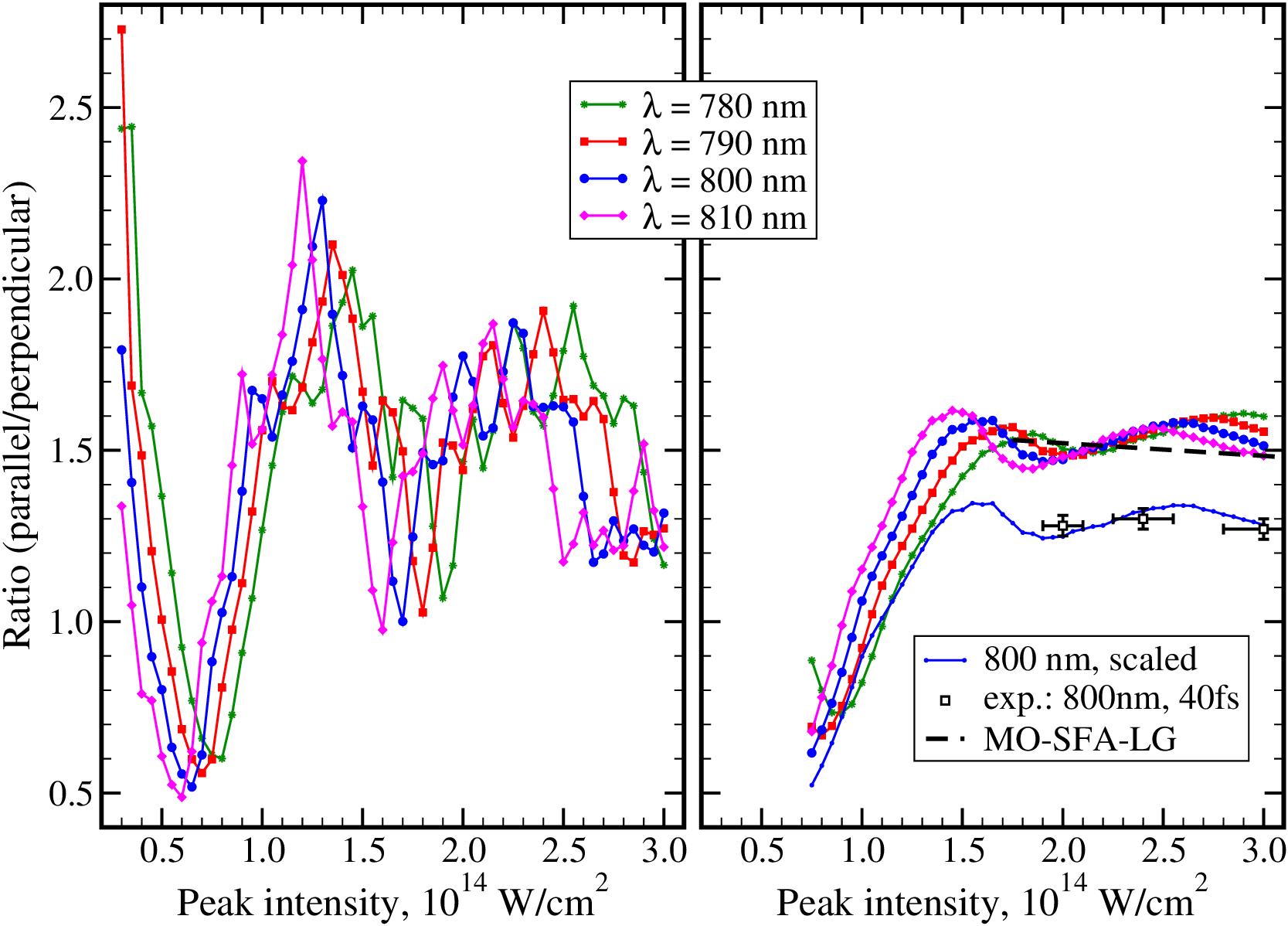}
     \caption{\label{fig:Ratios_IonY}
     (Color online) Ratio of parallel to perpendicular ionization yields
     of H$_2$ ($R=1.4\,a_0$) as a function of 
     laser peak intensity for different wavelengths and 10-cycle
     cos$^2$-shaped laser pulses: without (left panel) and with (right panel)
     focal-volume averaging. For a better comparison to the experimental 
     data from \cite{sfm:stau09} the 800\,nm result is also shown scaled 
     down by a factor 1.18. Furthermore, the (simplified) MO-SFA-LG data 
     in \cite{sfm:stau09} are depicted.}
\end{figure}

Figure~\ref{fig:Comp_IonY} shows the ionization yields for a parallel 
and a perpendicular orientation of H$_2$ 
exposed to a 800\,nm laser pulse as a function of laser peak intensity 
(for a fixed internuclear separation $R=1.4\,a_0$). As in all 
calculations shown in this work 10-cycle cos$^2$-shaped laser pulses
($\sim10\,$fs FWHM) 
are used. The numerical approach for solving the TDSE of H$_2$ has been 
described recently~\cite{sfm:vann08}. It is based on an expansion of the 
time-dependent wavefunction in terms of box-discretized field-free 
eigenstates obtained from a configuration-interaction calculation 
based on H$_2^+$ eigenstates calculated in a $B$-spline basis. 
According to Fig.~\ref{fig:Comp_IonY} the ionization yields show 
evident structures from resonantly-enhanced multiphoton ionization 
(REMPI) and channel closings, despite the rather high intensity and 
long wavelength. Since these structures depend on the orientation 
(the selection rules lead to different REMPI intermediate states and 
the different polarizabilities to different shifts of the multiphoton 
thresholds), the results for parallel and perpendicular orientation 
vary differently as a function of intensity. This is especially 
evident, if the ratio between parallel and perpendicular alignment 
is considered (Fig.~\ref{fig:Ratios_IonY}). This ratio shows a 
highly oscillating structure. Noteworthy, in a certain intensity 
range the perpendicular alignment leads to a larger ionization 
yield that the perpendicular one. A small variation of the wavelength 
within 780 and 810\,nm confirms that even the fine details of the 
structures are reproducible and not some numerical artifact. Although 
focal-volume averaging damps the sharp structures, the ratio remains 
non-monotonous and in some peak intensity regime the perpendicular 
orientation is easier to ionize than the parallel one.           

Very recently, the ratio of the ionization yields for parallel and 
perpendicular alignment was measured for 800\,nm laser pulses 
in~\cite{sfm:stau09}. The results are also shown in 
Fig.~\ref{fig:Ratios_IonY}. Note, that the intensities have been 
adopted as in~\cite{sfm:stau09}, although the experiment was 
performed with circular polarized light. This is in agreement to 
the analysis 
in~\cite{sfm:stau09} that compared to a model for linear polarization 
using unscaled intensities. Supported also by the present finding, 
the authors of~\cite{sfm:stau09} argued that due to experimental 
reasons the found anisotropy may be underestimated. Dividing the 
TDSE results by a factor 1.18 gives in fact a very satisfactory 
agreement between theory and experiment. Although not really 
statistically relevant in view of the error bars, one may note 
that also the experimental data indicate a small local maximum 
whose exact position depends according to the present TDSE results 
on the exact wavelength.   
In~\cite{sfm:stau09} a simplified MO-SFA-LG model is also proposed 
that considers basically the different polarizabilities of H$_2$ 
for parallel and perpendicular orientation. The model is in the 
intensity regime shown in~\cite{sfm:stau09} in good overall 
quantitative agreement with the present TDSE result, but does not 
show the small minimum. More importantly, it appears unlikely that the 
model could explain the sharp decrease and even inversion of the 
ratio for slightly smaller intensities. 

In the case of a homonuclear diatomic molecule with internuclear 
separation $R$ a bonding HOMO $\Phi$ 
built from s-type atomic orbitals $\phi$ is given within the LCAO 
as
\begin{equation}
  \Phi(\vek{r},\vek{R}) = a \{ \phi(\vek{r}, -\vek{R}/2) + 
                                    \phi(\vek{r},\vek{R}/2)  \}
\label{eq:LCAO}
\end{equation}
where $\vek{r}$ is the electronic coordinate and $a$ the normalization 
constant. According to MO-SFA-VG~\cite{sfm:muth00} (also called 
first-order IMST) this leads to the $N$-photon ionization rates 
(integrated over the directions of the emitted electron) 
\begin{equation}
 \Gamma_N = 
    N_e \int d\, \hatb{k}_{N} \frac{ d\,W^{(N)}}{d\,\hatb{k}_{N}}
\label{eq:GammaN}
\end{equation}
in a linearly polarized laser field, if the HOMO is occupied by $N_e$ 
electrons. The in Eq.~(\ref{eq:GammaN}) occurring differential 
ionization rates are given by
\begin{align}
\nonumber
\frac{ d\,W^{(N)}}{d\,\hatb{k}_{N}} &= 2\, \pi\, C^2\, k_N\, (U_p - N \omega)^2
J_N^2\Bigl(\bsymbol{\alpha}_0\cdot\vek{k}_N,\frac{U_p}{2\omega}\Bigr) \\
&\times \Bigl| 2 a \bk{\vek{k}_N}{\phi} \Bigr|^2 \cos^2( \vek{k}_N\cdot\vek{R}/2)  \; .
\label{eq:parGamma}
\end{align}
Here,  $k^2_N/2 = N\omega - (U_p + E_{\rm ion})$ is the kinetic energy of an
electron after absorption of $N$ photons, $U_p=F^2/(4\omega^2)$ is the
ponderomotive energy of an electron in a laser field of frequency $\omega$
and peak field strength $F$, $E_{\rm ion} = \kappa^2/2$ is the ionization 
energy of the molecule, and $C^2=(\kappa^3/F)^{2/\kappa}$ is a Coulomb 
correction factor. Finally, $\bk{\vek{k}}{\phi}$ 
is the Fourier transform of the atomic orbital $\phi(r)$ and $J_n(a,b)$ 
is a generalized Bessel function of two arguments as defined 
in \cite{sfa:beck05}. 
The polarization axis $\bsymbol{\varepsilon}$ enters the ionization 
rate only through one of the arguments of the Bessel function, 
$\bsymbol{\alpha}_0 = (F/\omega^2)\bsymbol{\varepsilon}$. 

The sum of the $N$-photon ionization rates for all energetically allowed 
values of $N$ yield the energy-resolved electron spectra also known as 
above-threshold ionization (ATI) spectra. Since the Fourier transform 
of spherically symmetric (s-type) orbitals is also spherically symmetric, 
it depends only on $k$. Therefore, the ratio of the $N$-photon ionization 
rates for parallel and perpendicular orientations of the field 
polarization vector with respect to the internuclear axis can be written 
as 
\begin{equation}
 X_N = \frac{\Gamma^{\parallel}_N}{\Gamma^{\perp}_N} = \frac{ 
 \int d\, \hatb{k}\,
 J_N^2\Bigl(g_N \bsymbol{\varepsilon}\cdot\hatb{k},b\Bigr) 
 \cos^2\Bigl(d_N \hatb{k}\cdot\hatb{R}^{\parallel}\Bigr) }{ 
 \int d\, \hatb{k}\,
 J_N^2\Bigl(g_N \bsymbol{\varepsilon}\cdot\hatb{k},b\Bigr) 
 \cos^2\Bigl(d_N \hatb{k}\cdot\hatb{R}^{\perp}\Bigr) }
\label{eq:ratio_XN_gen}
\end{equation}
where $g_N = \alpha_0\, k_N$, $b = U_p/(2\omega)$, $d_N = R\, k_N/2$, and 
all factors depending only on the absolute value of $\vek{k}_N$ were 
taken out of the integral and cancel each other when the ratio is 
considered. 

Fixing the coordinate system in such a way that its $z$ axis 
agrees with the polarization vector $\bsymbol{\varepsilon}$,  
one has $\hatb{R}^{\parallel} = (0,0,1)$, $\hatb{R}^{\perp} = (1,0,0)$, 
and $\hatb{k} = (\sin\theta\cos\phi,\sin\theta\sin\phi,\cos\theta)$. 
Then the ratio~(\ref{eq:ratio_XN_gen}) can be rewritten as
\begin{equation}
 X_N = \frac{ 
 \int_{0}^{\pi}d\theta\,\sin\theta\,  J_N^2\Bigl(g_N \cos\theta,b\Bigr) 
 \cos^2(d_N \cos\theta) }{ 
 \int_{0}^{\pi}d\theta\,\sin\theta\,  J_N^2\Bigl(g_N \cos\theta,b\Bigr) 
[1 + J_0(2 d_N \sin\theta) ]/2 }
\label{eq:ratio_ATI}
\end{equation}
where the identity 
\begin{equation}
 \int_0^{2\pi}d\phi \cos^2(\delta  \cos\phi) = \pi [ 1 + J_0(2\delta) ]
\label{}
\end{equation}
for zero-order Bessel function of the first kind was used. 

For very strong fields the function $J_N^2(g_N \cos\theta,b)$ 
peaks usually sharply around $\cos\theta = \pm 1$ (i.e.\ in the case of
ionization from a spherically symmetric atom the vast majority of 
electrons are ejected parallel to the laser polarization axis). 
On the other hand, the functions $\cos^2(d_N \cos\theta)$ and 
$J_0\bigl( 2 d_N \sin\theta \bigr )$ vary rather slowly in these 
regions. They may thus be approximated by their values at 
$\cos\theta \approx \pm 1$: 
\begin{equation}
\cos^2(d_N \cos\theta) \approx \cos^2(d_N),\qquad 
J_0\bigl( 2 d_N \sin\theta \bigr ) \approx 1 \; .
\label{eq:approx}
\end{equation}
Substitution of Eq.~(\ref{eq:approx}) into Eq.~(\ref{eq:ratio_ATI})
results finally in a very simple expression for $X_N$, 
\begin{equation}
 X_N \approx \cos^2(d_N) = \cos^2(R k_N/2) \; .
\label{eq:ratio_approx}
\end{equation}
Note, that for very small $d_N$ (which means small $R$ or small $k_N$) 
both functions in Eq.~(\ref{eq:approx}) are equal to 1, so the 
ratio $X_{N}$ is also 1. 

According to Eq.~(\ref{eq:ratio_approx}) MO-SFA-VG predicts a pronounced 
minimum to occur at the electron energy $E \approx \pi^2/(2R^2)$, if the 
ratio between the the energy-resolved electron spectra obtained 
for a parallel and a perpendicular orientation is considered. Its origin 
is a pure interference phenomenon that is due to the destructive 
interference caused by the phase shift between two electronic wavepackets 
emerging from the two nuclei and moving along the polarization axis.

\begin{figure}
\begin{center}
     \includegraphics[width=0.44\textwidth]{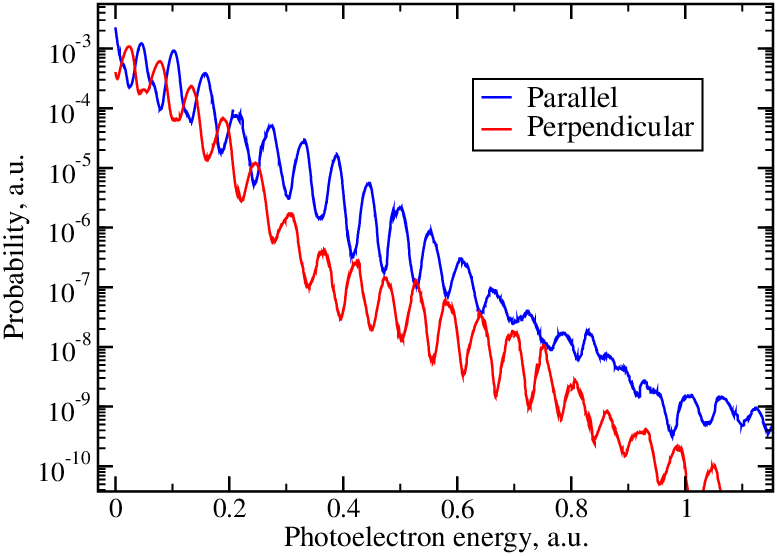}
     \caption{\label{fig:Comp_ATI}
     (Color online) Energy-resolved electron spectra for parallel and 
      perpendicular orientations of H$_2$ ($R=3.0\,a_0$) 
      for a 10-cycle cos$^2$-shaped 800\,nm laser
      pulse with a peak intensity of $2\cdot10^{13}\,$W/cm$^2$.}
\end{center}
\end{figure}

\begin{figure}
\begin{center}
     \includegraphics[width=0.44\textwidth]{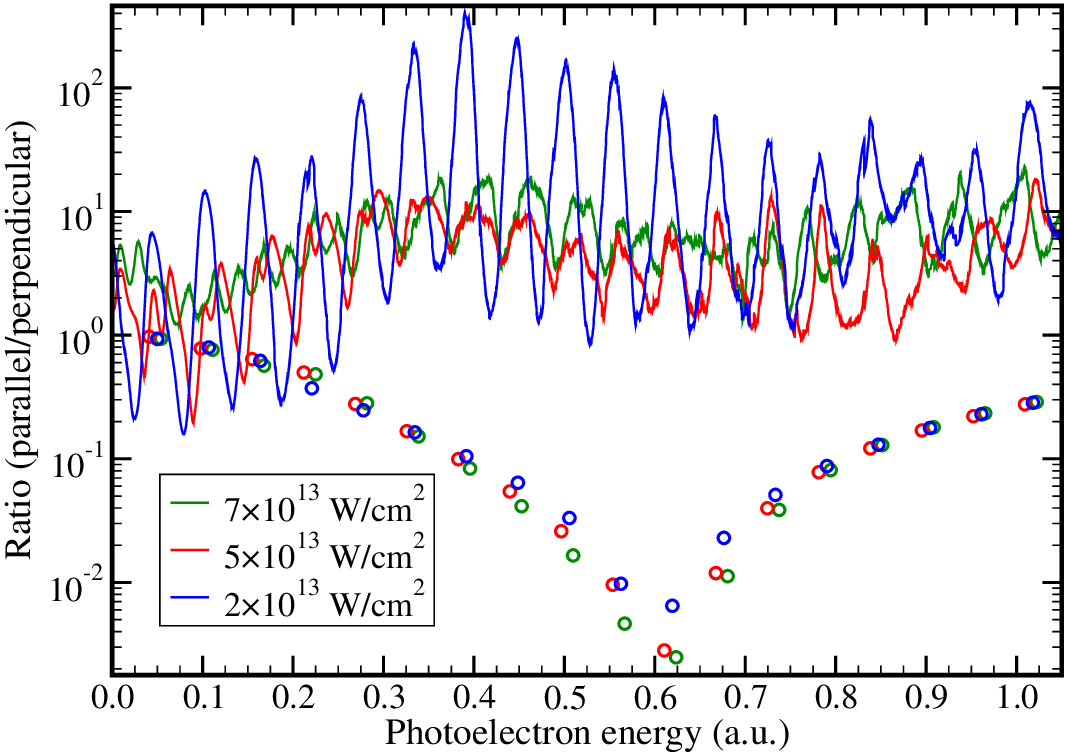}
     \caption{\label{fig:Ratios_ATI}
     (Color online) Ratio of parallel to perpendicular 
      energy-resolved electron spectra of H$_2$ ($R=3.0\,a_0$) 
      for 10-cycle cos$^2$-shaped 800\,nm laser
      pulse and three different peak intensities. 
      The ratios predicted by MO-SFA-VG in 
      Eq.~\ref{eq:ratio_ATI} are also shown using circles.}
\end{center}
\end{figure}

In order to investigate the occurrence of this interference phenomenon 
predicted by MO-SFA-VG, the energy resolved electron spectra of H$_2$ 
were also extracted from the TDSE calculations. In order to amplify 
the effect, the larger internuclear separation $R=3.0\,a_0$ is 
considered in Fig.~\ref{fig:Comp_ATI}. This should be an ideal case, 
since at these distances the HOMO of H$_2$ is extremely well described 
by two hydrogenic 1s orbitals. There are, in fact two more reasons for 
choosing a larger $R$ value. If the predicted minimum lies at too 
high energies, rescattering that is not incorporated in the first-order 
SFA theory could dominate and cover the interference phenomenon. 
Furthermore, the energy-resolved spectra obtained from the TDSE 
calculation converge more easily for lower energies, due to their  
fast exponential decay. Despite all these attempts to find perfect 
conditions for finding the interference minimum predicted by MO-SFA-VG, 
the ratio of the electron spectra for the two orientations does not 
show the behavior predicted by MO-SFA-VG in Eq.~(\ref{eq:ratio_ATI}) 
(that agrees well 
to the approximate expression (\ref{eq:ratio_approx})) as can be 
seen from Fig.~\ref{fig:Ratios_ATI}. It is interesting that the TDSE 
results indicate in fact a rather universal overall behavior of the 
ratio independent of the laser intensity. However, the ratio 
first increases with increasing electron energy and remains on average 
above unity. As is clear from Fig.~\ref{fig:Comp_ATI}, the total 
ionization yield stems dominantly from the very low-energy 
electrons. In this regime the ratio between the electron spectra 
for parallel and perpendicular is, however, rather intensity dependent. 
This explains the already discussed intensity dependence of the total 
yield found already at the equilibrium distance.

\begin{acknowledgments}
The authors acknowledge financial support from the {\it Stifterverband 
f\"ur die Deutsche Wissenschaft}, the {\it Fonds der Chemischen Industrie},  
and {\it Deutsche Forschungsgemeinschaft} (Sa\,936/2). 
\end{acknowledgments}

\bibliographystyle{apsrev}

\end{document}